\documentclass[conference]{IEEEtran}
\IEEEoverridecommandlockouts

\usepackage{cite}

\ifCLASSINFOpdf
  \usepackage[pdftex]{graphicx}
  \DeclareGraphicsExtensions{.pdf,.jpeg,.png}
\else
  \usepackage[dvips]{graphicx}
  \DeclareGraphicsExtensions{.eps}
\fi

\usepackage[cmex10]{amsmath}
\usepackage{array}
\usepackage{fixltx2e}
\usepackage{url}

\def\E{{\cal{E}}}
\def\para#1{{\bf{#1}}}

\begin{document}
%
\title{A First Practical Fully Homomorphic Crypto-Processor Design\\[1ex]
\Large The Secret Computer is Nearly Here}

\author{\IEEEauthorblockN{Peter Breuer}
        \IEEEauthorblockA{Email: Peter.T.Breuer@gmail.com}
\and
        \IEEEauthorblockN{Jonathan Bowen}
        \IEEEauthorblockA{London South Bank University\\
           London, UK\\
           Email: jpbowen@gmail.com
        }
}

\maketitle

\begin{abstract}
Following a sequence of hardware designs for a fully homomorphic
crypto-processor -- a general purpose processor that natively runs
encrypted machine code on encrypted data in registers and
memory, resulting in encrypted machine states -- proposed by the authors
in 2014, we discuss a working prototype of the first of those, a
so-called `pseudo-homomorphic' design.  This processor is in principle
safe against physical or software-based attacks by the owner/operator of
the processor on user processes running in it.  The processor is
intended as a more secure option for those emerging computing paradigms
that require trust to be placed in computations carried out in remote
locations or overseen by untrusted operators.

The prototype has a single-pipeline superscalar architecture that runs
OpenRISC standard machine code in two distinct modes.  The processor
runs in the encrypted mode (the unprivileged, `user' mode, with a long
pipeline) at 60-70\% of the speed in the unencrypted mode (the
privileged, `supervisor' mode, with a short pipeline), emitting a
completed encrypted instruction every 1.67-1.8 cycles on average in real
trials.

\end{abstract}


%
\IEEEpeerreviewmaketitle

\section{Introduction}
\IEEEPARstart
In 2013, the authors published theory \cite{BB13a} showing that if the
arithmetic in a standard processor is modified appropriately, then,
given three provisos detailed below, the processor continues to operate
correctly, but all the states obtained in the processor and memory are
encryptions of the states obtained in an unmodified processor running
the same program.  That is a rather unintuitive result, given the common
experience that modifying a computer program even slightly, or even a
slight bug in an arithmetic library -- or worse, in a hardware unit --
gives rise to catastrophically different program results.  Nevertheless,
it is so, as the 2013 paper proved, and a second paper in 2014
\cite{BB14b} set out several routes to implemention.  However, the
provisos referred to above are not trivial, and represent engineering
challenges to be overcome on the way to a practical product.  In this
paper we report on a completed prototype that implements one of the
first and simplest of the options discussed in \cite{BB14b}, which,
following the nomenclature there, will be referred to as KPU (`krypto
processor unit') designs.  The KPU designs work by modifying arithmetic,
and the outcome is a processor in which user mode processes run with
encrypted data in memory, in registers, and on busses.  In principle,
the operator physically in charge of the processor can see only
encrypted data with physical probes, while supervisor mode processes,
which run unencrypted within the processor, likewise can see only
encrypted user data with software probes.  The KPU therefore has
relevance to running securely in the Cloud, or for tamper-proofing
computing machinery such as voting or banking machines in physically
remote locations.

\para{The  first proviso} is that {\em the modified
arithmetic within the processor of a KPU must be a `homomorphic image'
of ordinary computer arithmetic}.  That does not imply that within the
processor merely some 1-to-1 rearrangement of the conventional (2s
complement) encoding of numbers in 32-bit binary has been effected.
In principle, an arbitrary 1-to-n encryption, such as Rijndael-64
\cite{DR2002}
with 32 bits of padding under the encryption, may be chosen.
The block size of the encryption $\E$ (64 bits for Rijndael-64) must
coincide with the register size and physical memory word size in the
processor, so using a 256-bit encryption means designing for
256-bit registers and memory busses, etc, but apart from that the choice
of encryption is up to the designer.

However, the `homomorphic image' property boils down to
mathematical constraints $\E(x+y)=f(\E(x),\E(y))$ and other relations
set out in \cite{BB13a}, which specify what the encrypted output
$\E(x+y)$ from the modified arithmetic logic unit (ALU) in the processor
must be when the encrypted inputs $\E(x)$ and $\E(y)$ are presented, so
the designer will choose an encryption that achieves a satisfactory
trade-off between the security of $\E$ and the feasibility of
implementing an appropriate function $f$ in hardware.  We will return to
this topic below, but we note at this point that the requirement 
is formally weaker than classical expressions of
homomorphism in encryption, which have $\E(x+y)=\E(x)+\E(y)$.
That is not quite a question of `what is in a name' as to whether 
the function is called `$f$' or `$+$'; it emphasizes that the
designer is not obliged to co-opt the familiar `$+$' for $f$.

\para{The second proviso} has to do with memory addressing in the KPU and the
kind of programs that can run in it.  Because data addresses look no
different from other numbers, and are produced dynamically in the course
of a program, for example by adding an offset to a base address, in the
KPU designs of \cite{BB14b}, {\em data addresses are
encrypted} exactly as other data is.
However, {\em program addresses are not encrypted} because the program counter
in any processor is most often advanced by a constant (the length of an
instruction in bytes) at each tick of the clock.  That would
allow an attack against the encryption, if the same encryption were used
for program addresses as for data addresses.  The simplest solution,
adopted by all the KPU designs in \cite{BB14b}, is not to encrypt
program addresses at all.

In consequence, {\em running programs must never combine program
addresses with ordinary data values}.  A conforming program may not
perform arithmetic computations on program addresses.  In other words,
it must never jump to a subroutine whose address is the square root of
Elvis's birthdate written backwards.  Programs that are written to
respect the distinction between the two types are called {\em
crypto-safe} in \cite{BB12a}, where a formal type-system for machine
code is set out that ensures machine code programs may run successfully
in a KPU.

That appears restrictive because, for example, dynamic loaders and
linkers compute program addresses at run time and they will not be able
to.  However, the restriction only applies to a KPU running in the
encrypted mode of operation, and the designs in \cite{BB14b} envisage
that it runs encrypted in the (unpriviliged) `user' mode while the
(privileged) `supervisor' mode runs unencrypted.  That allows dynamic
loaders to operate successfully in the conventional way in supervisor
mode.

\begin{figure}[!t]
\centering
\includegraphics[width=2.25in]{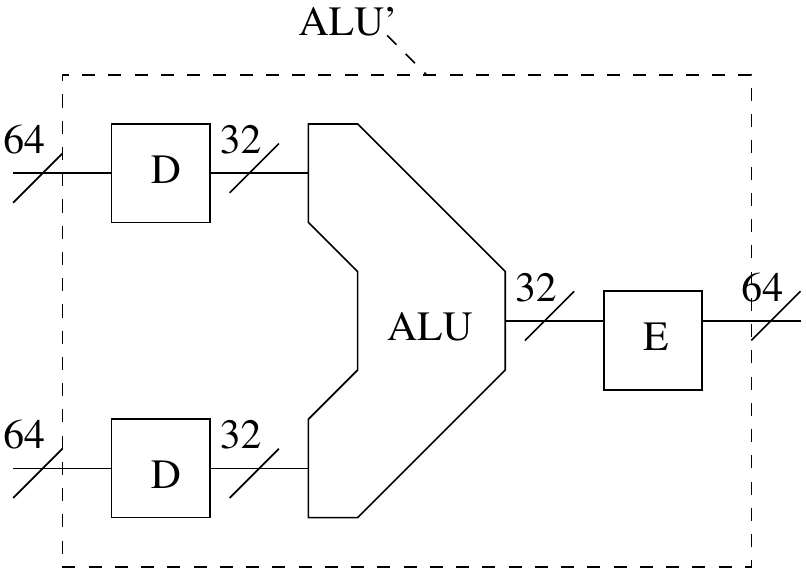}
\caption{A 32-bit processor arithmetic logic unit (ALU) modified for
encrypted operation (ALU$'$) with a 64 bit block size by the addition of
decryption units (D) on the inputs and an encryption unit (E) on the
output.}
\label{f:1}
\end{figure}

\para{The third proviso} referred to in the first paragraph of this
article is due to the fact that many different binary codes may be
generated during execution for what the programmer intended to be
the same memory address, as a consequence of the encryption of memory
addresses and the 1-to-n nature of good encryptions.  All the KPU
designs in \cite{BB14b} treat memory as a possible adversary that should
not be privy to the encryption, so there is no way for the memory unit
to know that these different encryptions should all alias the same data.
That gives rise in programming terms to 
{\em hardware aliasing}, in that the same address (as seen by a
program running under the encryption) sporadically accesses
different data.  To avoid it, {\em programs have to be compiled
following a particular style} \cite{BB14a,BB14c,BBP15}.  The `trick'
depends on the processor being deterministic at bottom
in all things, including calculation of the encrypted
addresses.  So to reproduce a particular encrypted address exactly it
suffices to store it for recovery later, or to repeat exactly the
calculation that produced it in the first place.  Making use of those
two coding strategies, compilation for the KPU works (we have modified
the GNU gcc 4.9.1 compiler and assembler port for the OpenRISC 1.1
architecture
(\url{opencores.org/or1k/Architecture_Specification})
to suit; the source code is at
\url{sf.net/p/or1k64kpu-gcc} and \url{sf.net/p/or1k64kpu-binutils}
respectively).

Of the KPU design options set out in \cite{BB14b}, the three that are
relevant to our prototype are:

\paragraph{Pseudo-homomorphic \label{o:a}} Instead of using an
encryption with special properties to achieve the necessary homomorphic
property (proviso \#1 above), or innovative hardware in the ALU, this is
in concept an ordinary RISC \cite{Pat85} design with the standard ALU
augmented only by encryption/decryption (`codec') units affixed inline
to the inputs and the outputs (Fig.~\ref{f:1}).  That is an elegant
design only from the mathematical point of view -- we demonstrated
formally in \cite{BB13a} that that design must work correctly, and in a
pilot project in Java (see \url{sf.net/p/kpu}) at the same time we
constructed an object model of a standard pipelined RISC processor and
dropped in an ALU modified with codecs `fore and aft' as in
Fig.~\ref{f:1} and verified experimentally that it operated as
predicted.  But the design (i) does not immediately offer speed (the
encryption and decryption of data on every instruction lengthens the
time to complete an instruction by a factor of twenty), (ii) does not
appear to offer good prospects for physical security, since unencrypted
data is being processed internally and encryption and decryption using
keys is taking place.

\begin{figure}[!t]
\centering
\includegraphics[width=1.8in]{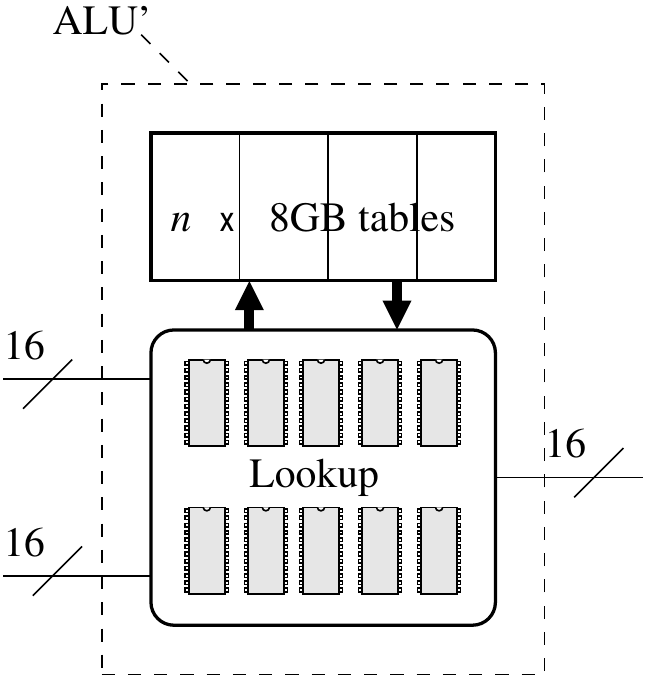}
\caption{A 16-bit encrypted arithmetic logic unit (ALU$'$) embedding 
8GB arithmetic tables for each of $n$ operators.}
\label{f:2}
\end{figure}

\paragraph{Lookup}
This design embeds the encrypted arithmetic tables for the ALU
directly (Fig.~\ref{f:2}).  That obviates both
the need for codecs and for keys (and corresponding questions of
provision and safekeeping).  However, the space requirement for as
little as 16-bit operation is already considerable, and 20-bit operation
is probably the limit with present technology (requiring access to
Terabyte-sized RAM storage for each of the tables for operations such as
addition or multiplication; note, however, that such motherboards
are already available off the shelf -- see for example the
Supermicro Xeon 7000 -based range at
\url{supermicro.com/products/motherboard/Xeon7000/}).  One might
consider increasing the number of bits by putting ALUs together in
modular fashion, but aside from anything else that means memory hardware
that offers multiple addressing for simultaneous retrieval, if the
storage requirement is not to multiply up with the number of modules.
At most dual-port (read twice simultaneously) RAM is available at
present, though quad-port RAM is not far off.

\paragraph{Partially homomorphic}
The problem with the fully homomorphic encryption (FHE; homomorphic
with respect to both addition and multiplication) first created by
Gentry \cite{Gentry09, cryptoeprint:2009:616} in 2009 and its subsequent
improvements and improved implementations at IBM and elsewhere
\cite{Halevi11, Brakerski:2012} is that it runs at on the order of one
bit-op per second, with approximately a million-bit block size, so it is
quite impractical for use.  Otherwise it would fit perfectly in a KPU
design, requiring no modification of the ALU from standard as far as
addition and multiplication go (division, shift right and the
comparator relations would still require modification, however).

A half-way house towards an encryption that achieves the `homomorphic
image' requirement on its own account is a partially homomorphic
encryption (PHE).  RSA \cite{RSA} is the canonical example, but it is
homomorphic with respect to multiplication, in that
$\E(x*y)=\E(x)*\E(y)$, which is not very convenient.  The Pallier
encryption \cite{Pal} is additively homomorphic instead, which lends
itself much more easily to use in a KPU.  The idea would be to implement
addition in the modified ALU as standard addition over the Pallier
encryption, making use of its homomorphic property,
$\E(x+y)=\E(x)+\E(y)$, to implement the other standard ALU
operations, multiplication, division, etc, in terms of addition.

Unfortunately, that is mathematically impossible.  However, a single
extra look-up table for the signs of encrypted numbers makes it
possible.  The explanation is that the single operation `${\bf
if}~(x{>}k_0 \mathop{\&} y>k_1 \mathop{\&} \dots)~{\bf
then}~x{{-}{=}}K_1,y{{-}{=} }K_2,\dots$' is computationally complete --
this is the only statement form in Conway's famous Fractran programming
language \cite{conway87fractran} --, and that may be implemented given the
homomorphic addition plus table of signs, then all the other 
(encrypted) operations of the ALU may be implemented in terms of it,
either directly in hardware or as software routines.  Indeed, an
encrypted processor based on a stack machine model (instead of the von
Neumann model) with the Fractran statement as its only non-control
machine instruction, together with a 16-bit Pallier crypto-system plus
table of signs is described in \cite{heroic}, so we have confirmation
that the Pallier system plus table of signs is feasible.  A
functional difference between the KPUs discussed here and the stack
machine of \cite{heroic} is that the latter does not experience hardware
aliasing (see proviso \#3 above), since no (encrypted) memory addresses
are involved (stack `addresses' are explicit unencrypted offsets from
the bottom of the stack, essentially individual local variable names).
However, stack machines are not generally implemented directly in
hardware, and it would be unwise to expect that interfacing one with
all the i/o interrupts, caches, busses and other paraphernalia found on
board a modern processor would be easy.  A 1 Terabyte table of signs is
enough to feed a 43-bit Pallier encryption, however, no matter whether
the encrypted processor is a stack machine or von Neumann
design.

Moreover, there is a technique set out in \cite{BB14b} that in principle
`cubes the security', while only tripling the storage requirement.  The
idea (`ABC typing') is to encrypt each of the elements $f(a,b)=c$ of an
encrypted calculation using a different encryption.  That is
$a=\E_A(x)$, $b=\E_B(y)$, $c=\E_C(z)$.  The compiler can generate code
that respects this type discipline and the processor ALU can be designed
to implement it.  While an attacker may guess the key for encryption A,
say, in order to know he/she is right about A, the keys for encryptions
B and C must also be guessed.  Only then can the attacker confirm that
an observed $c$ encrypts $z$ that is the sum of $x$ and $y$.  If the
keyspace for each of encryptions A, B and C is size 43 bits, then
$3\times43=129$ bits of keyspace must be searched overall.  The
technique is readily applied when the function $f$ is implemented via
lookup table, requiring three lookup tables instead of one (inputs of
type A and B, B and C, C and A respectively are valid).  In the case
when the $f$ is the Pallier system addition, since the Pallier partial
homomorphism may be `upgraded' to fully homomorphic by a lookup table of
signs (the homomorphic multiplication then being implemented by a
software routine) the technique described by Gentry in \cite{Gentry09}
for changing from one FHE to another without decryption may be applied
(essentially: run the decryption algorithm of one over the other's
encryption, using the second's encryption of the first's key).

Another technique discussed in \cite{BB14b} and applicable to
all design solutions is to encrypt each bit of data differently.  A
16-bit lookup-table solution for a 1-bit processor becomes a 32x16-bit
lookup-table solution for 32-bit operation, requiring also 16-to-16-bit
translation tables between 1-bit ALU modules.  But the technique does
not improve security beyond 16 bits, because the arithmetic of the least
significant bit can be attacked independently, then the second bit can
be attacked, and so on, so we will not consider that direction here.

\para{The prototype described in this paper} implements option (a)
above, the `pseudo-homomorphic' option.  Option (b), a lookup
table-based solution, is not viable at present, because too much memory
is required for acceptable security.  Oprion (c), using the Pallier or
similar partially homomorphic cryptosystem, also still needs a
Terabyte-size table of signs for even 43 bit encryption, which is also
not sufficiently secure for general use.  The enhancements using ABC
typing discussed above would improve the numbers, but it is still the
subject of research.  To be acceptable for, say, banking applications,
right now a pseudo-homomorphic solution using a standard $n$-bit
encryption best fits the bill.  Moreover, $n$ can conveniently be
adjusted from one prototype to another in line with hardware resources,
matching register size and bus width.

The objections to a pseudo-homomorphic solution are those already noted:
(i) it is slow, (ii) it is physically vulnerable.  However, (ii) is an
objection already overcome  \cite{SmartCard} by Smartcard manufacturers,
who overlay parts of their chips that contain keys and encryption
apparatus with delicate traces that cause the chip to fail if disrupted
by a physical probe, so it is not an insuperable
objection.  Memory does not contain unencrypted user data in a KPU, so
it is not vulnerable to `cold boot' \cite{Simmons:2011, Gruhn2013,
halderman2009lest} attacks either (essentially, physically freezing the
memory sticks in order to retain an image of the DRAM contents even
without power), and only the processor chip needs Smartcard-like
protection.  As to (i), we have been able to innovate in the
architecture design so as to achieve in encrypted running 60-70\% of the
speed in unencrypted running, with an encrypted instruction being
completed every 1.67-1.8 cycles, as described below.

\section{Architecture}

The prototype KPU is based on the OpenRISC 1.1 32/64 bit architecture
and instruction set specification
(\url{opencores.org/or1k/Architecture_Specification}).  It runs
instructions uniformly 32 bits in length on 32- or 64-bit data stored in
memory and registers.  Encrypted data physically occupies 64 bits, but
it contains only 32 bits of meaningful data when decrypted.

In user mode, the processor runs on encrypted data and executes only
the 32-bit instruction set (i.e., those instructions that target 32-bit
data).  A 64-bit instruction run in user mode raises an `illegal
instruction' exception.  As per the OpenRISC specification, user mode
instructions access all 32 general purpose registers (GPRs), and also a
very few permitted special purpose registers (SPRs).  Attempts to write
`out of bounds' SPRs are ignored in user mode and zero is read.

In supervisor mode the processor may execute either 32- or 64-bit
instructions and access to registers is unrestricted.  There is no
enforced division of memory into `supervisor' and `user' parts, so
a supervisor mode process can read user data from memory, but
the user data will be in encrypted form.

OpenRISC instructions divide into two kinds: `immediate' instructions,
which carry 16 bits of data in the (32-bit) instruction itself, and
`register' instructions, which do not.  The immediate instructions are
problematic in user mode because we want them to carry data in
encrypted form.  But encrypted data takes up 64 bits and an instruction
is only 32 bits long, so it does not fit.  To solve this problem, a {\em
prefix} instruction has been added to the instuction set.  An immediate
instruction will be preceded in the instruction stream by two prefix
instructions, each carrying a 24-bit segment of the encrypted datum, and
the immediate instruction itself carries only the final 16-bit segment.
Those OpenRISC immediate instructions that are supposed to carry fewer
than 16 bits of data (register shifts and rotations each carry 5 or 6
bits) have been respecified to contain exactly 16 bits of data.

\begin{figure}[!t]
\centering
\includegraphics[width=3.5in]{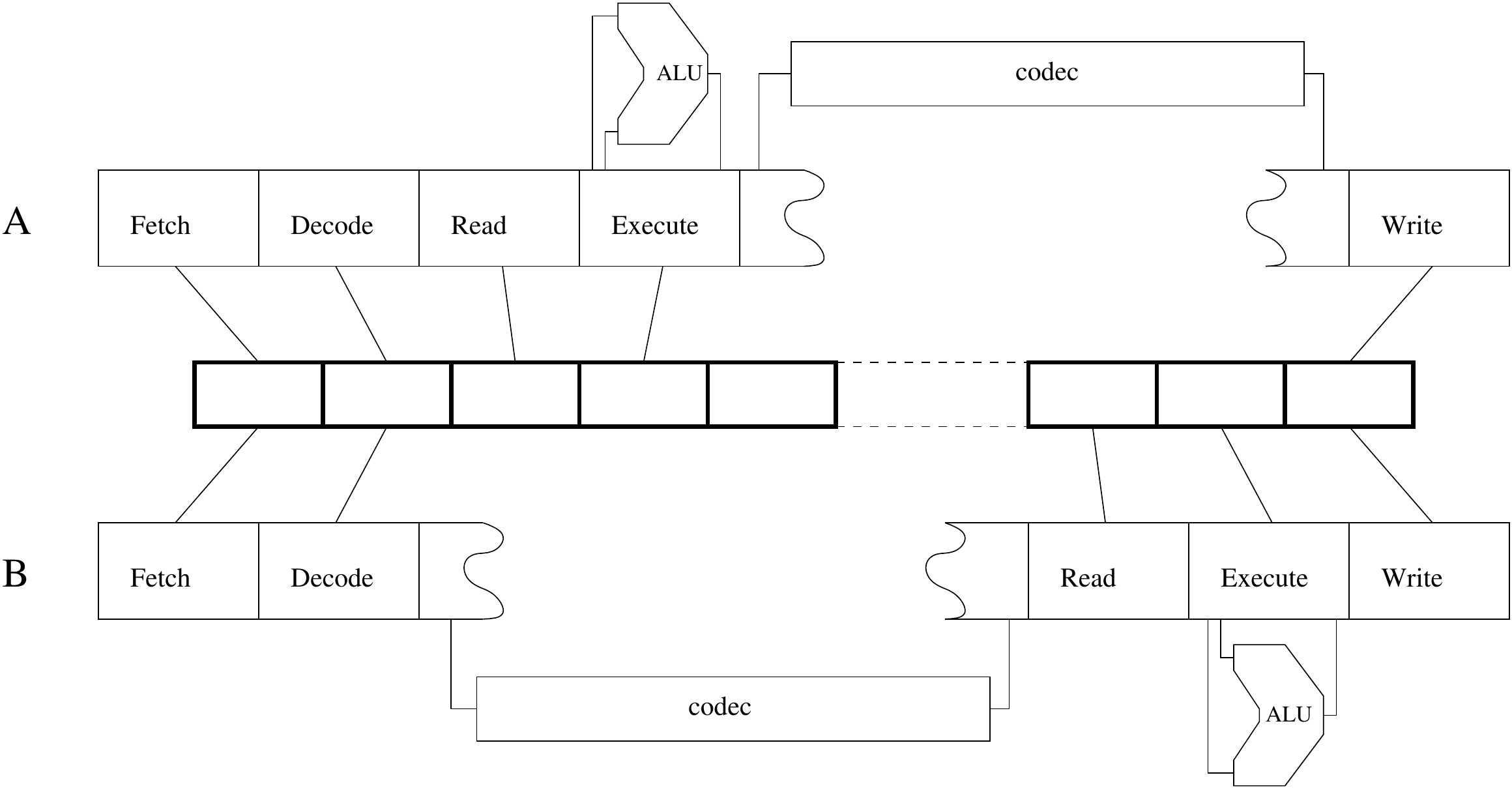}
\caption{Physically the same pipeline is configured in two different 
ways, `A' and `B', for two different kinds of user mode instructions
during encrypted working.}
\label{f:3}
\end{figure}

The instruction pipeline in (unencrypted) supervisor mode is the
standard short 5-stage fetch, decode, read, execute, write pipeline 
expected of a RISC processor \cite{Pat85}, except that it is physically
embedded in a longer pipeline that is traversed in full by (encrypted)
user mode instructions.  The pipeline is configured in two different
ways for the user mode instructions as shown in Fig.~\ref{f:3} (the
hardware for those stages with two different configurations is
doubled).  The reason is that, in order to reduce the frequency with
which codecs are brought into action for user mode instructions, ALU
operation is effectively {\em extended in the time dimension}, so that
it covers a series of consecutive (encrypted) arithmetic operations in
user mode.  Only the beginning of the series is associated with a
decryption event, when encrypted data in memory or registers is converted, and
only the end of the series is associated with an encryption event.
Longer series mean less frequent codec use.  It turns out that two
pipeline configurations cover the needs of instruction processing when
codec use is required.

The `A' configuration is deployed when a store instruction puts an
encrypted result into memory, or a load instruction decrypts incoming
data from memory.  The `B' configuration is used when encrypted
immediate data in an `add immediate' instruction is read in.
Instructions that do not exercise the codec pass through with the
pipeline in `A' configuration, because the early execution makes results
available for early forwarding to instructions entering behind, avoiding
pipeline stalls.  The codec covers 10 stages in this implementation.

To support this mode of operation, the ALU posseses a private set of
user-mode-only registers that shadow the GPRs (and the few SPRs
accessible in user mode).  These are intended to contain the decrypted
version of the encrypted data in the `real' GPRs and SPRs.  They are
mapped in during read and write stage of a user mode instruction, and
mapped out for supervisor mode instructions, so they are unavailable to
supervisor mode.  This protocol automatically maintains the register
entries in decrypted form in the shadow registers from one instruction
to the next during user mode operation.

Additionally, a small user-mode-only data cache caches the unencrypted
version of any encrypted data that is written to memory duing user mode
operation.  On load from memory, this cache is checked first.  Almost
all execution stack reads in normal operation are intercepted by this
mechanism.  The cache is physically within the processor boundary, so
will be covered by the measures that protect the processor chip from
spying or interference (e.g., Smartcard-like fabrication).

Note that in the KPU, program addresses are unencrypted (as opposed to
data addresses, which are encrypted), which potentially is a source
of confusion in user mode, because unencrypted data is kept in shadow
registers within the ALU, while encrypted data is passed to memory
and the `real' registers. A particular protocol addresses the issue:
unencrypted 32-bit addresses zero-filled to 64 bits are regarded as the
`encrypted' form, and they are `decrypted' to an `unencrypted' form
consisting of the same data with the top 16 bits of 64 rewritten to
0x7fff.  Thus an instruction such as jump-and-link (JAL) in user mode,
which fills the return address (RA) register with the program address of
the next instruction, writes the zero-filled address to the RA shadow
register, and the 0x7ff form to the real RA register. The padding
under the encryption is always arranged so that real encrypted data
avoids looking like either of these forms of program address.

In principle, encrypted addresses emanating from the KPU fall anywhere in
the full 64-bit range (although the addresses under the encryption are
32-bit).  Since no real machine ever has a full 64 bits-worth of memory
available, conventionally address translation takes place
within the memory management unit to a physically backed area of memory
via a `translation look-aside buffer' (TLB).  However, the TLB is
conventionally organised at page-sized granularities, saying where each
8KB-sized area of logical addressing should be translated to in physical
addressing terms.  That architecture is not appropriate for a KPU,
because encrypted addresses are not clustered, if the encryption
is any good.  Instead, the KPU's TLB must be organised with word-sized
granularity.  Further, all encrypted addresses generated in user mode
are remapped by the TLB to a pre-set range with the allocation serially
ordered by `first-come, first-served'.  Since data that will later be
accessed together tends also to be addressed for the first time in close
sequence, this allows conventional cache lookahead policies to operate
successfully.

Moreover, it has turned out to be possible in this `pseudo-homomorphic'
design to pass the unencrypted data address to the memory unit
during the processing of load and store instructions, with no additional
processing.  We are nervous of the security implications, so we do not
suggest that that should be done.  However, the bare 32-bit address
could be hashed or encrypted in a different way to 64 bits from there.

\section{Performance}

The OpenRISC `or1ksim' simulator
(\url{opencores.org/or1k/Or1ksim}) has been modified to run the
KPU prototype discussed here.  The code comes with the OpenRISC
specification-compliant exception conditions and actions built in, as
well as a comprehensive set of peripheral devices, caches, buffers, etc.
It contains a Verilog compiler backend as an alternative to direct
execution in the simulator.  The simulator was upgraded to 64 bit
simulation from 32 bits (see \url{sf.net/p/or1ksim64ptb}), following the
OpenRISC 1.1 specification document at
\url{opencores.org/or1k/Architecture_Specification#OpenRISC_1000_architecture_1.1},
and then the processor core simulation was changed to encode
the pipeline discussed in this paper, with full forwarding of data
between pipeline stages for instructions running in the same processor
mode, and a branch prediction cache.  Performance measurements on the
simulator are made directly on the pipeline, and are not estimates.
Supervisor mode and user mode are accounted separately, so comparisons
between encrypted and unencrypted working are made on the same
architecture.  That modified simulator code is available from
\url{sf.net/p/or1ksim64ptb}.

The processor instruction set tests from the or1ksim suite have been
modified to run in a KPU.  The original tests ran in supervisor mode,
which would not have tested a KPU, in which supervisor mode is
unencrypted.  The assembler source code was rewritten to run in
(encrypted) user mode instead, making system calls when access to data
such as the state of the carry or overflow flags in the processor status
register (SR) is required for the test reports.  OpenRISC's port of the
GNU `gas' assembler v2.24.51 has been modified to produce encrypted
machine code for this KPU target and the modified source code is
available at \url{sf.net/p/or1k64kpu-binutils/}.

One question that modifying the testsuite has settled is whether or not
it is possible to run encrypted code that does anything useful in a
conventional sense.  After all, local peripherals work unencrypted,
without privy access to encryption mechanisms in the processor or
elsewhere, so even output on the local display seems a priori to be
problematic.  Fortunately, the original or1ksim code is `tricked out'
with a special debugging no-op instruction that sidesteps the need to
interact with an output peripheral and the testsuite uses it to do all
its report I/O.  The `trick' no-op prints the contents of register r3
directly on the screen, and we have not had to do anything more for I/O
other than maintain the use of those no-op instructions in the testsuite
code.  In user mode, the ALU's shadow r3 register provides a decrypted
value for printout, and in supervisor mode the value in the r3 register
is unencrypted, so the testsuite printout is comprehsible to an
observer.

Aside from printing, however, yes, it has been possible to write all the
(thirty or forty) system calls for execution in supervisor mode that
proved necessary.  There are calls, for example, to blindly set or clear
various SR flags that determine if an
arithmetic overflow triggers an exception or not.  System status flags
that may be set in user mode are cleared by the hardware when
changing to supervisor mode, so supervisor mode cannot get information
from them (the flags are saved by the hardware in a special register
to which supervisor mode has no access for recovery to user mode later),
but that has not proved fatal to full reporting of the exception
handling.

\begin{table}[!t]
\renewcommand{\arraystretch}{1.3}
\caption{Performance data, or1ksim testsuite instruction set add test.}
\label{t:1}
\centering
\begin{tabular}{|r||r|r|}
\hline
\multicolumn{3}{|l|}{@exit  : cycles 315640, instructions 222006}\\
\hline
\hline
            mode  &user        &super     \\
\hline
 register  instructions  & 0.2\%      &  0.2\%   \\
 immediate instructions  & 7.3\%      &  9.2\%   \\
 load      instructions  & 0.9\%      &  2.8\%   \\[-1ex]
          (cached)& ( 0.9\%)   &          \\
 store     instructions  & 0.9\%      &  0.0\%   \\[-1ex]
          (cached)& ( 0.9\%)   &          \\
 branch    instructions  & 1.0\%      &  4.9\%   \\
 jump      instructions  & 1.1\%      &  4.8\%   \\
 no-op     instructions  & 6.4\%      & 15.8\%   \\
 prefix    instructions  &11.5\%      &  0.0\%   \\
 mf/tspr   instructions  & 0.1\%      &  2.7\%   \\
 sys/trap  instructions  & 0.5\%      &  0.0\%   \\
 wait     states  &24.7\%      &  4.9\%   \\[-1ex]
          (stalls)& (22.1\%)   &  ( 3.8\%)\\[-1ex]
         (refills)& ( 2.7\%)   &  ( 1.1\%)\\
\hline
           total  &54.8\%      & 45.2\%   \\
\hline
\end{tabular}

\medskip
\begin{tabular}{|r|r||r|r|}
\hline
\multicolumn{4}{|c|}{Branch Prediction Buffer}       \\
\hline
      hits   &10328     ( 55\%)& misses   &8219  ( 44\%) \\
\hline
      right  &8335      ( 44\%)& right    &6495  ( 35\%) \\
      wrong  &1993      ( 10\%)& wrong    &1724  ( ~9\%) \\
\hline
\hline
\multicolumn{4}{|c|}{User Data Cache}    \\
\hline
read  hits &2942 (99\%)& misses &0 (~0\%)   \\
write hits &2933 (99\%)& misses &9 (~0\%)   \\
\hline
\end{tabular}
\end{table}

Table~\ref{t:1} displays the performance statistics summary from the 
modified instruction set add test (`is-add-test') of the or1ksim
testsuite.  The statically compiled executable contains 185628 machine
code instructions, which occupy 742512 bytes in the 769454 byte
executable, the rest being comprised of the executable file headers,
symbol table, etc.  Table~\ref{t:1} shows that when this test was run
(successfully) to completion, 222006 instructions were executed, so
there are few loops and subroutines (the code is largely built using
assembler macros) in 315640 cycles.  If one reckons with a 1GHz clock,
then the speed was just over 700Kips (instructions per second) overall.

In supervisor mode, pipeline occupation is just under 90\%, at 892Kips
for a 1GHz clock (wait states, cycles in which the pipeline fails to
complete an instruction, comprise 4.9\% of the 45.2\% total), which one
may take as a baseline for a single pipeline superscalar design.  In
user mode pipeline occupation is only 54.9\%, as measured by numbers of
non-wait states, for 549Kips with a 1GHz clock.  Measured against
supervisor mode, that is 61.6\% of the unencrypted speed.

The wait states are caused by real data hazards in the pipeline.  Most
(84\%) are due to a load instruction feeding directly to an arithmetic
instruction.  The stall occurs because the data address for the load
instruction is only calculated in execute stage, so the data cannot
at that time already be available to the instruction sitting in read
stage just behind.

The data indicates that a dual pipeline might be beneficial, perhaps
enabling speed over 70\% of unencrypted running.

\section{Conclusion}
A superscalar pipeline design for an encrypted processor has been
described here, with performance measured at 60-70\% of unencrypted
processing while embedding a 10-cycle (Rijndael) 64-bit encryption.
Registers, memory and busses contain encrypted data in this
`pseudo-homomorphic' design.


%
%



\bibliographystyle{IEEEtran}
\bibliography{IEEEabrv,icics2016}
%
%
%

\end{document}